\documentclass[APS,twocolumn,floatfix]{revtex4}
\usepackage{graphicx}
\usepackage{calc}
\usepackage{pifont}
\usepackage{amssymb}
\usepackage{epsfig}
\usepackage{amssymb}
\usepackage{psfig}
\usepackage{amsmath}
\begin{document}

\title{Differential Conductance Peak Exchange in Borromean Ring\\
Single Molecule Transistors}

\author
{Gavin D. Scott$^{1}$, Kelly S. Chichak$^{2}$, Andrea J. Peters$^{2}$, Stuart J. Cantrill$^{2}$, J. Fraser Stoddart$^{2}$, and H. W. Jiang$^{1} $}

\affiliation{$ ^{1}$Department of Physics and Astronomy, University of California, Los Angeles, Los Angles, California 90024\\
             $ ^{2}$Department of Chemistry and Biochemisty, University of California, Los Angeles, Los Angles, California 90024}
\begin{abstract}
While Single Molecule Transistors have been employed in laboratory research for a number of years, there remain many details related to charge transport that have yet to be delineated.  We have used the technique of electromigration to create a nanometer-size gap for hosting a single molecule in the narrowest portion of a gold wire defined by electron beam lithography.  A recently synthesized Borromean Ring complex was used as the active molecular element of the transistor devices.  It is well established in Single Molecule Transistors that, for sweeps of source-drain voltage, a region of zero conductance commonly appears - known as a conductance gap - which is centered about $V_{sd} = 0$.  We have observed differential conductance peaks on opposing sides of these conductance gaps exchange positions as a result of changing gate bias.  We associate this property with a change in sign of the majority charge carriers flowing through the molecule.  That is, a measured change in differential conductance occurs when current through a molecular energy level shifts between the flow of holes and the flow of electrons.
\end{abstract}

\maketitle

\textbf{INTRODUCTION} The field of molecular electronics has grown dramatically in recent years with its increasing feasibility in addition to its potential for new limits of device scaling.  The premise of electronic devices with characteristics dictated by molecules, both small and large, has been in existence for several decades.  While some of the earliest work involving electrical transport measurements through organic molecules was performed in the laboratory of Hans Kuhn\cite{Mann1971} in the early 1970s, it wasn't until the 1990s that experiments were performed with the ability to contact individual molecules and nanocrystals.\cite{Reed1997,Klein1997,Klein1996} It was at this time that the field of molecular electronics began to blossom.  Using a single molecule as an active element represents an extreme in the miniaturization of modern electronic devices, and is one of the contemporary goals of molecular electronics research.  To this end, our efforts have been focused on the challenge of gaining a greater understanding of electrical transport through such nanoscale devices.  As our comprehension of the electrical properties of molecules grows, so does our ability to incorporate molecules into electrical devices\cite{Heath2003,Flood2004} such as rectifiers, interconnects, photovoltaics, crossbar memories, and biosensors. 

Single Molecule Transistors (SMTs) have proven to be innovative devices for investigating transport through isolated molecules.  In fact, the device has the potential to be useful in studying many sorts of nanoscale structures.  The SMTs we fabricate are composed of a source and a drain electrode plus a conducting substrate that is used as a back gate.  We employed the electromigration breakjunction technique first described by Park \textit{et al.}\cite{Park1999} The molecular compound we chose to use was a Borromean Ring  (BR) complex (Figure 1a-c), the synthesis of which was described recently.\cite{Chichak2004}

On account of the extreme sensitivity of the device, measuring the conductance through a metal-molecule-metal junction can not only tell us about the Coulomb charging energy of the system, but it can also give us information about individual molecular orbital levels present in the molecule, as well as vibrational modes of the molecule and correlated electron states.\cite{Yu2004}  This is why the SMT is such a unique tool for studying physical phenomena on the nanometer scale.  We have produced stability diagrams that display differential junction conductance as a function of both gate voltage and source-drain voltage.  These stability diagrams illustrate a phenomenon in which two distinct differential conductance peaks on opposing sides of the conductance gap move relative to each other as a function of the applied gate voltage, thereby increasing or decreasing the size of the conductance gap.  For a given gate bias, a plot of differential conductance versus source-drain voltage is often markedly asymmetric.  This characteristic allows one to distinguish between the two differential conductance peaks.  As the gate voltage is decreased from zero, the two peaks approach one another, as they both advance toward zero source-drain voltage, until they coalesce and ultimately close the conductance gap.  If the gate voltage is decreased even more, the conductance gap reopens, and the peaks recede from one another.  However, we observe that when the conductance gap reopens, the differential conductance peaks have exchanged positions.  A plot of the peaks after the exchange reveals that modulating the gate voltage has resulted in a reflection of the peaks about the differential conductance axis.  We believe that this conduction phenomenon is a consequence of a transition between current resulting from the flow of holes and current resulting from the flow of electrons.  Details of SMT transport, such as those presented here, lead us toward a larger picture of conduction through single molecules and nanoscale devices in general, as well as toward the eventual development of integrated molecular electronic devices.

\textbf{EXPERIMENTAL} Fabrication of the SMTs begins with the preparation of arrays of devices on an $n^+ Si$ substrate with a 140 $\AA$ $SiO_2$ insulating layer.  1.5 nm of Ti and 15 nm of Au are evaporated onto nanowire patterns defined by electron beam lithography.  Bowtie-shaped constriction patterns are produced with minimum widths of less than 100 nm.  After the bare devices have been finished, additional gold pads are defined using UV lithography, followed by the deposition of 10 nm Ti and 200 nm Au.  This fabrication procedure allows us to make electrical contact to the source and drain electrodes of the samples through wire-bonding.  The array of samples is exposed to an oxygen plasma for 1 minute to clean the surface of residual organics.  Following the method of applying molecules onto a substrate described by Yu and Natelson,\cite{Yu2004} a dilute solution of the BR complex (1 mg/2 ml) is subsequently spin cast onto the array of devices at 900 rpm for 30 seconds.  For this experiment the zinc-containing BR molecules (Figure 1a-c) were used.  The devices are then cooled to 4.2 K in liquid helium, after which the electromigration technique is used to produce a nanometer-size gap, located at the narrowest portion of the patterns (Figure 1d).  This approach entails ramping the DC source-drain voltage through the gold wires until a sharp drop in the conductance is observed.  The assumption is made that if a BR complex is immediately on top of the narrowest portion of a gold wire sample prior to breaking, then after the electromigration process is complete, the molecule will reside in the newly formed gap (Figure 1e).  A configuration will result such that electrons will be able to tunnel sequentially off a gold electrode and onto the molecule, and then from the molecule onto the second gold electrode.  

Molecular BRs have presented a formidable challenge to synthetic chemists in search of the realization of molecules with a high degree of topological sophistication.  The BR complex consists of three interlocking molecular rings, composed of macrocyclic organic ligands interacting with six zinc(II) ions (Figure 1a-c).  The complex was constructed by multiple cooperative self-assembly processes from eighteen original components - six endo-tridentate ligands, six exo-bidentate ligands, and six transition metal ions.

\begin{figure}[htb]
\scalebox{1}{\includegraphics[width=8.5cm,height=10.6cm,angle=0]{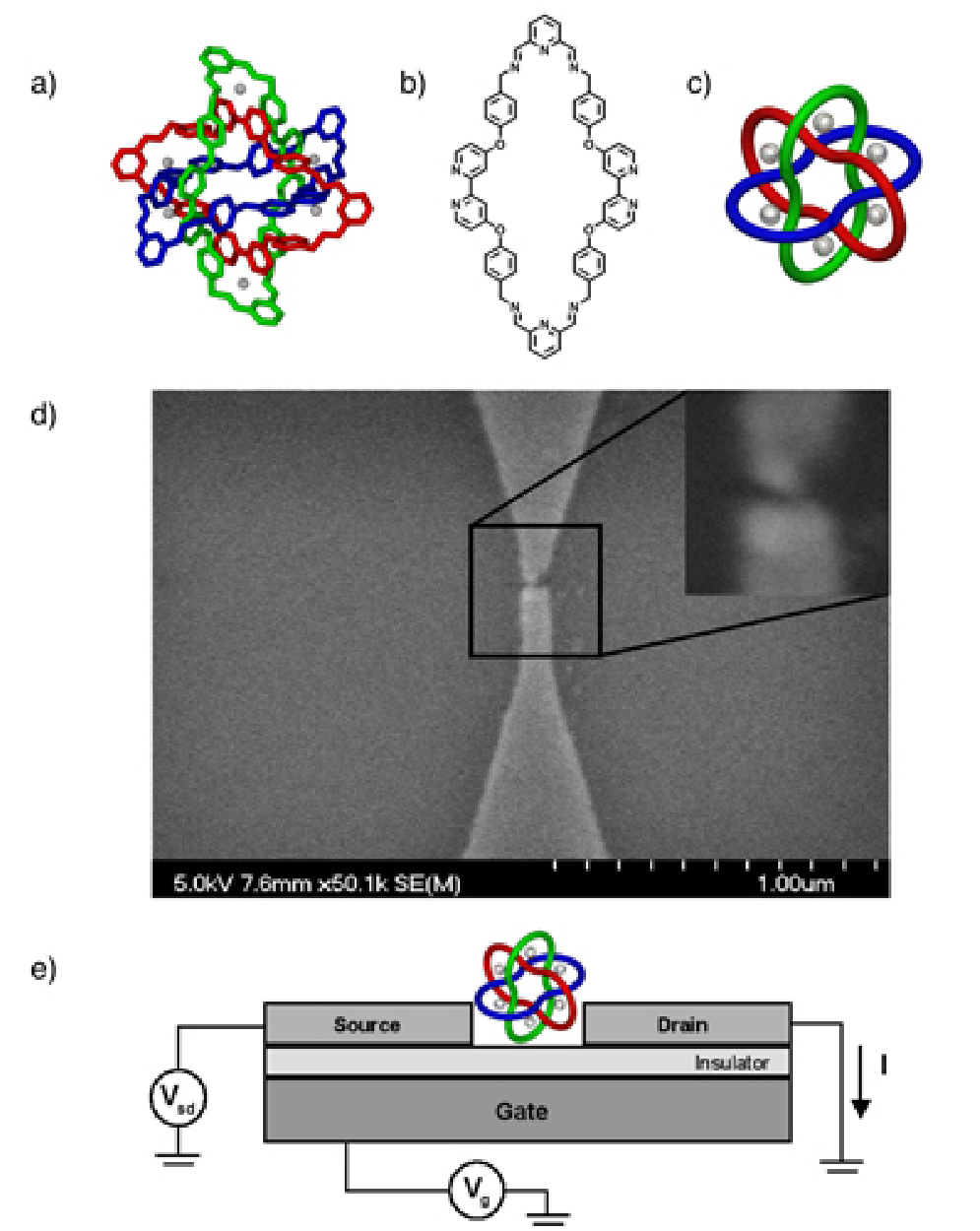}}
\caption{a) The X-ray crystal structure of the zinc(II)-containing Borromean Ring complex. b) The constitution of the macrocyclic ligand which is repeated three times over in the Borromean Ring complex. c) A graphical representation of the Borromean Ring complex. d) SEM picture of breakjunction after electromigration. e) Schematic representation of an SMT with the desired molecule-electrode configuration. }
\end{figure}

The BR complex was selected as the active element of the SMTs for a variety of reasons.  First of all, these molecules are approximately 2.5 nm in diameter, making them ideally suited for use in conjunction with the breakjunction method of SMT fabrication.  When electromigration causes a gap to form in the gold wire, the width of the gap is typically on the order of a few nanometers.  Since the average gap is significantly wider than most molecules, larger molecules will more easily 'bridge' the gap and thus be easier to contact electrically.  Another advantage of the BR complex is that it is shaped more or less like a sphere, and thus it is roughly symmetric along its x, y, and z-axes.  With this lack of anisotropy we hope to alleviate inconsistencies that may arise when measuring conductance through its various axes, since we currently have no way of controlling the precise orientation of the molecule once it is situated within a breakjunction.  By contrast, a long thin molecule may show different conductance behavior if current flows through its long axis as compared to its short axis.

One of the most significant reasons for working with BR complexes is the ability we have to choose the transition metal ions.  The BR complex is a relatively large molecule in which we can locate during synthesis six transition metals, be they Mn(II), Fe(II), Co(II), Ni(II), Cu(II), and Zn(II), or whatever.  Thus, we have the ability to repeat experiments with the same basic ligand structure while only altering the transition metal ions.  In the future, this diversity will allow us to investigate the relationship between a small portion of the molecule's constituents and a device's transport properties. 

\textbf{RESULTS} Traces of differential conductance ($\partial{I}/\partial{V}$) were recorded as a function of source-drain voltage, $V_{sd}$.  These graphs were obtained at incrementally changing values of the gate voltage, $V_g$.  When $\partial{I}/\partial{V}$ is plotted as a function of both $V_{sd}$ and $V_g$, we form a stability diagram in which black represents regions of zero conductance and brighter areas are indicative of increased conductance (Figure 2).  A myriad of physical properties are revealed in the stability diagram of an SMT.  The most obvious feature is the changing conductance gap centered around zero source-drain voltage.  The conductance gap arises from a combination of the Coulomb Blockade effect, due to the small size and correspondingly large charging energy of the molecule, and to the discrete energy level spacing of the molecular orbital levels.  The stability diagram in Figure 2 also shows indications of electrons accessing excited states of the molecule.  We attribute this to the Fermi levels of the electrodes aligning with vibrational resonance modes of the molecule.\cite{Park2000,Pasupathy2005}  For a given gate voltage, a graph of $\partial{I}/\partial{V}$ as a function of $V_{sd}$ will typically appear asymmetric, with two distinctly different $\partial{I}/\partial{V}$ peaks on opposing sides of the conductance gap (Figure 3).  This is due to the asymmetric coupling of electrodes and the charge-induced effects that result.\cite{Zahid2004}  When scanning $V_{sd}$, the first conductance peaks to appear on opposing sides of the conductance gap - call them the 'central peaks' - are usually of different height and width and do not necessarily appear at identical values of $\mid V_{sd}\mid$.  A prominent feature of the present data is the exchange of position of these two peaks about zero $V_{sd}$, a feature that is evident as a function of changing $V_g$. 

\begin{figure}[htb]
\scalebox{1}{\includegraphics[width=8cm,height=6.5 cm,angle=0]{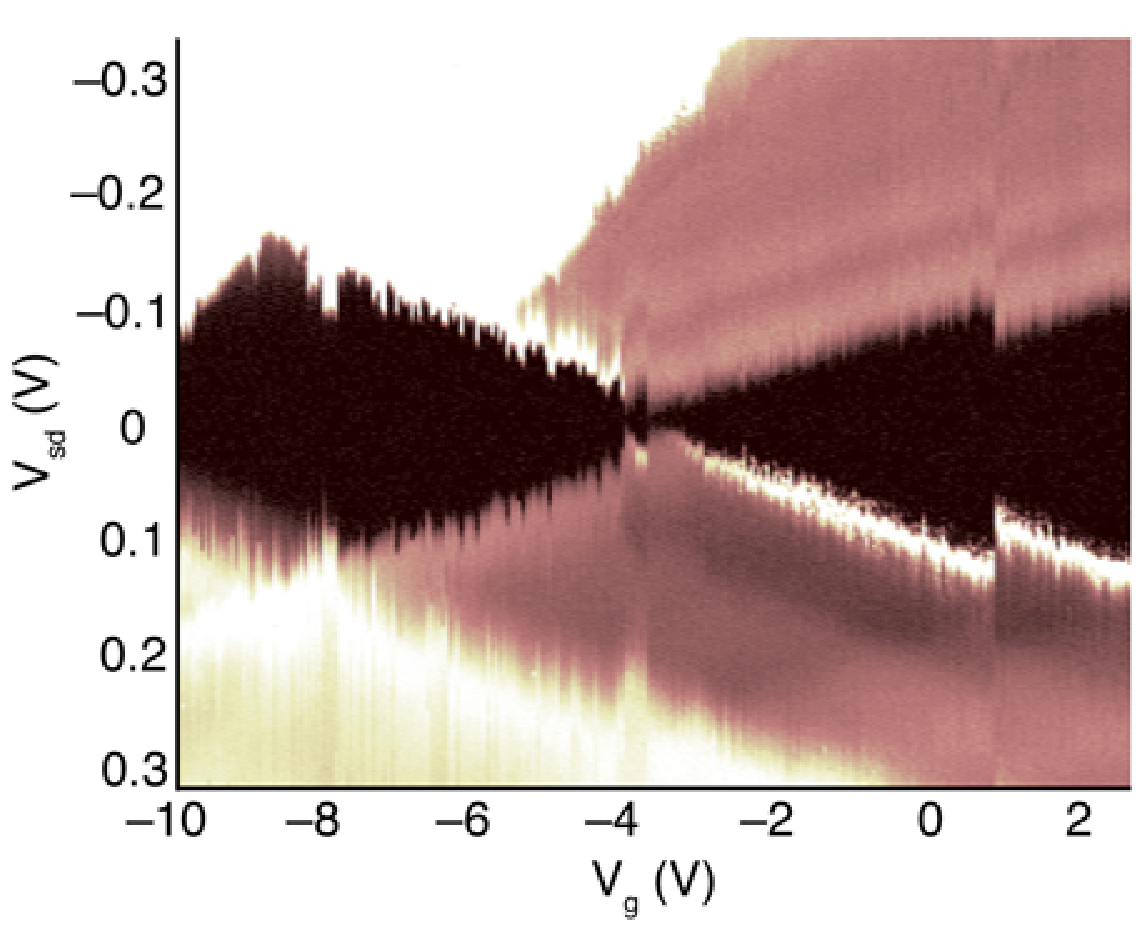}}
\caption{Differential conductance ($\partial{I}/\partial{V}$) measured as a function of both source-drain voltage ($V_{sd}$) and gate voltage ($V_g$).  Black denotes zero conductance, and white denotes the maximum conductance.  Sidebands on upper right side indicate electrons accessing excited states.}
\end{figure} 

A peak in $\partial{I}/\partial{V}$, which would be represented as a step in an I-V trace, occurs on the edge of the conductance gap when the source and drain electrodes have been biased such that the Coulomb charging energy has been overcome and the Fermi level of one electrode first aligns with an energy level of the molecule.  The $V_{sd}$ traces of Figure 3 were taken at successively increasing values of $V_g$.  We observe that, as $V_g$ is increased, the two central peaks begin to move toward one another (Figure 3b-e).  In other words, in this energy range, as the voltage from the gate decreases the energy of the molecule, the conductance gap shrinks.  After the conductance gap closes, at voltage $V_g = V_C$, it then begins to reopen as $V_g$ continues to be increased, and we see the central peaks begin to separate (Figure 3f-i).  This observation is in agreement with the Coulomb Blockade effect as it pertains to the flow of current through microscopic structures.  What we also see is that the two distinct central peaks have switched sides with respect to $V_{sd} = 0$.  Rather than the two conductance peaks approaching one another until the conductance gap closes, and then simply receding from one another past $V_g = V_C$, what we see instead is that the two distinct peaks approach one another until $V_g = V_C$, and then appear to move past one another.  In this sense we continue to see an asymmetric graph of $\partial{I}/\partial{V}$ versus $V_{sd}$, however there has been a reversal of the asymmetry resulting in a reflection of the differential conductance graph about the $\partial{I}/\partial{V}$ axis.  This peak interchange has been observed for all of our working SMT samples.  Although there is evidence of such conductance peak exchanges in data published by Park \textit{et al.},\cite{Park2003} and while this feature of a stability diagram has been observed before in other SMTs, to date there has been no discussion of its origins.  It should be noted that not all conductance peaks evident in a given SMT behave similarly.  While the central peaks always exchange position, some peaks we have observed are apparently unaffected by the application of gate voltage, and other peaks move in a direction opposite to that of the central peaks.  Clearly, not all $\partial{I}/\partial{V}$ peaks are produced as a result of the exact same process. 

\begin{figure*}[t!]
\scalebox{1}{\includegraphics[width=16.2cm,height=12.5 cm,angle=0]{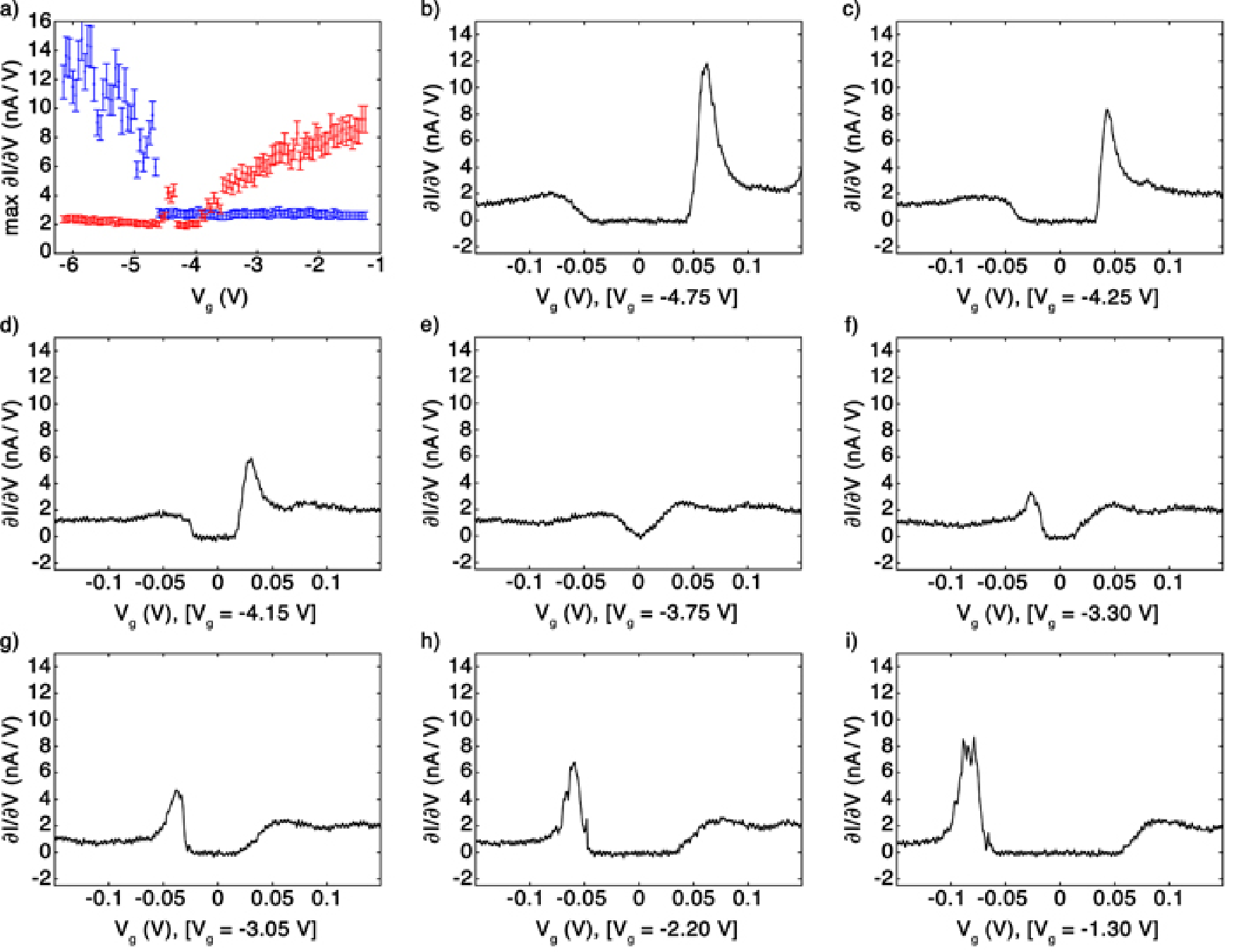}}
\caption{a) Graph plotting ($\partial{I}/\partial{V}$) peak heights as a function of applied gate voltage.  Blue data points correspond to ($\partial{I}/\partial{V}$) peaks for $V_{sd} > 0$, and red data points correspond to ($\partial{I}/\partial{V}$) peaks for $V_{sd} < 0$.  (b-i) Traces of differential conductance ($\partial{I}/\partial{V}$) taken at successively increasing gate voltages ($V_g$). For $V_g > V_C$ a positive source-drain bias produces a larger conductance peak than a negative source-drain bias (b-e).  As $V_g$ is increased, the central peaks move toward 0$V_{sd}$ until $V_g = V_C$, after which they exchange places.  For $V_g < V_C$ a positive source-drain bias produces a smaller conductance peak than a negative bias (f-i). }
\end{figure*}

\textbf{DISCUSSION} In equilibrium, the source and drain electrodes and the molecule all have a common electrochemical potential, $\mu$, the Fermi level.  When the source and drain electrodes are biased such that the source electrode has chemical potential $\mu_1$ and the drain electrode has chemical potential $\mu_2$, current will flow as long as an energy level of the molecule lies between $\mu_1$ and $\mu_2$ (at finite temperatures this range is extended by the thermal energy\cite{Ghosh2002} $\pm$$k_B$T).  This condition holds whether the energy level is filled or unfilled.\cite{Datta2004}  In the simplest case we will assume that neither the source or drain electrode is coupled too tightly to the molecular orbital levels.  As a source-drain bias is applied, $\mu_1$ and $\mu_2$ are displaced from $\mu$ by approximately equal and opposite amounts, $\pm\frac{1}{2}eV$.  For the case of an unfilled energy level, an electron will first pass from the electrode with higher chemical potential to the molecule.  The electron will then pass to the electrode with a lower chemical potential.  This process is continually repeated, establishing a steady state flow of electrons resulting in n-type conduction.  This will be the case for either a positive or a negative bias of the source and drain.  For the case of a filled energy level lying between $\mu_1$ and $\mu_2$, an electron will initially pass from the filled energy level of the molecule to the electrode with a lower chemical potential, leaving a hole in the filled energy level.  Subsequently, this hole will be replenished by an electron passing from the electrode with a higher chemical potential to the molecular energy level.  Again, a current is established, but this time it is a consequence of the steady state flow of holes resulting in p-type conduction.  So, we see that current can flow through a filled or unfilled molecular orbital level because it is not a question of the number of electrons on the molecule, but rather the number and relative positions of molecular orbital levels near the equilibrium chemical potential of the electrodes.  For an identical number of energy levels that charge carriers can flow through, conduction due to electrons and holes may differ on account of their differing effective mass and band structure.

In the SMTs the source and drain electrodes are not symmetrically coupled to the molecule.  The origin of the asymmetric $\partial{I}/\partial{V}$ graph can be understood by considering conduction occurring through the HOMO (highest occupied molecular orbital) level when one electrode is weakly coupled to the molecule while the other electrode is strongly coupled.  A positive bias on the weakly coupled electrode will cause electrons from the molecule to empty out to that electrode, while the strongly coupled electrode will move electrons onto the molecule, immediately replenishing any empty states.  Conversely, a negative bias on the weakly coupled electrode will cause electrons on the molecule to empty out to the strongly coupled electrode first, however they will not be replenished by the weakly coupled electrode as quickly.  This delay in response causes a net positive charge to build up on the molecule.  The positive charge lowers the energy level and delays the point at which current can begin to flow.  This delay is manifested as a stretching out of one side of the I-V graph, a phenomenon which then results in a reduced $\partial{I}/\partial{V}$ peak.  For conduction occurring through the LUMO (lowest unoccupied molecular orbital) level, the argument is reversed since a negative charge will be added to the molecule.  The end result is that a differential conductance peak will be lower for positive bias on the strongly coupled contact for hole conduction, and higher for electron conduction.  Therefore, plots of $\partial{I}/\partial{V}$ versus $V_{sd}$ for p-type and n-type conduction will be reflections of one another about the $\partial{I}/\partial{V}$ axis.  Figures 2, 3, and 4 show that a more pronounced differential conductance peak appears on the $V_{sd} > 0$ side of the graphs for n-type conduction, and on the $V_{sd} < 0$ side for p-type conduction.  Knowing that there must be a positive bias on the strongly coupled contact for p-type conduction, and that the polarity of $V_{sd}$ is taken relative to the source electrode, we conclude that, for the SMT used to obtain the data in Figures 2 and 3, the source electrode is strongly coupled to the molecule and the drain is weakly coupled.  When looking at a plot of $\partial{I}/\partial{V}$ versus $V_{sd}$, it is necessary to determine whether conduction is due to the flow of holes or electrons before concluding which electrode is more strongly coupled to the molecule.  Similarly, it is an indication that the majority charge carrier has switched between electrons and holes when the asymmetric nature of the differential conductance graph is reversed.  In this way it is the asymmetry that allows us to observe the conductance switch between n-type and p-type as it is evinced in the stability diagram.  If a relatively symmetric SMT were produced, conductance would still switch between n-type and p-type as a function of $V_g$, however it would not be as apparent.

\begin{figure}[htb]
\scalebox{1}{\includegraphics[width=6.85cm,height=14.9 cm,angle=0]{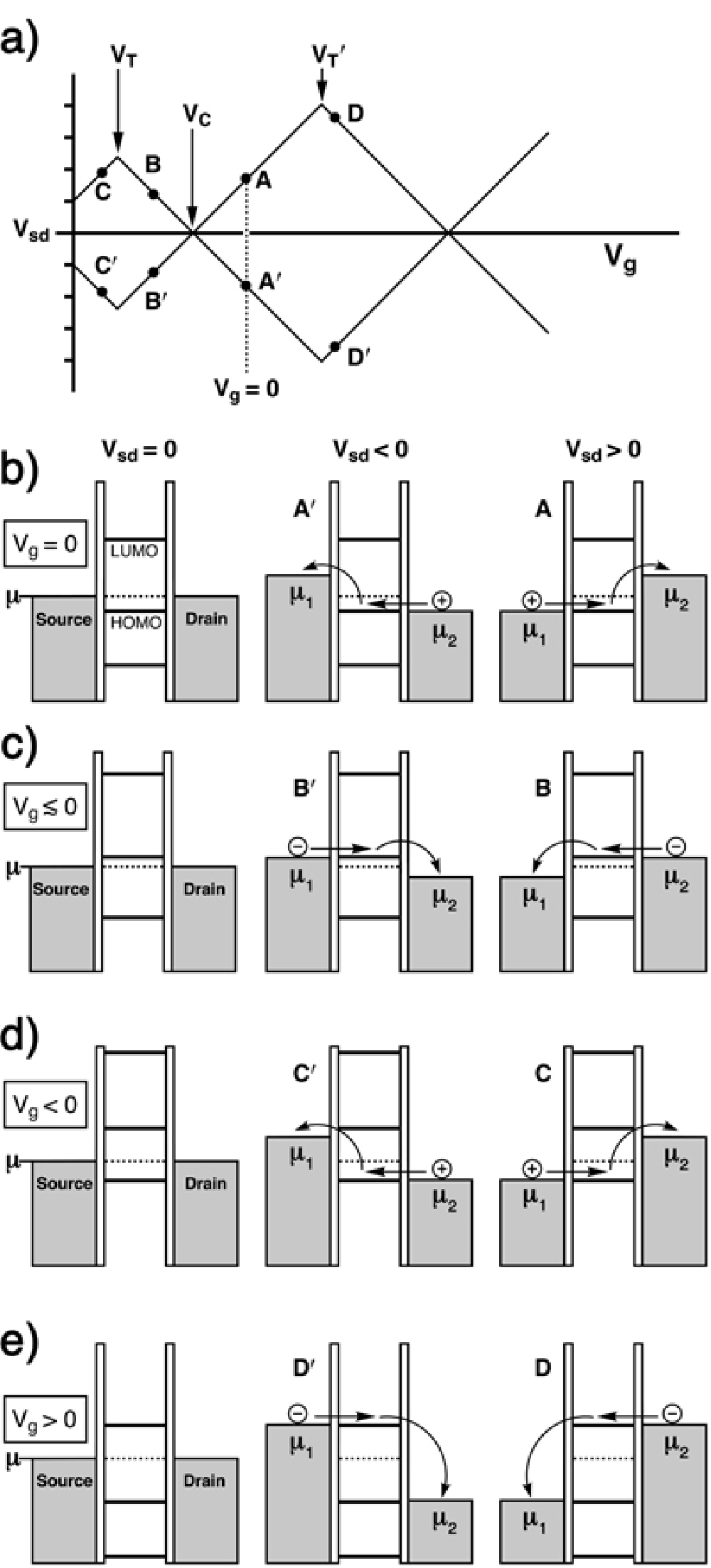}}
\caption{a) Schematic illustration of a stability diagram.  The black line indicates points where conductance becomes non-zero.  b) Energy level diagram of electrodes and BR complex for $V_g = 0$.  p-type conduction occurs through filled energy level for both positive and negative source-drain biases  c) Energy level diagram of electrodes and BR complex for $V_g < 0$ such that the electrode Fermi level is slightly below HOMO level.  n-type conduction occurs through empty energy level for both positive and negative source-drain biases.   d) Energy level diagram of electrodes and BR complex for $V_g < 0$ such that Fermi level is more than half way to next energy level down.  p-type conduction results.   e) Energy level diagram of electrodes and BR complex for $V_g > 0$ such that Fermi level is now closer to HOMO level.  n-type conduction results. }
\end{figure}

By viewing the interchange of differential conductance peaks to be indicative of a switch between p-type and n-type conduction, we can analyze a prototypical system to gain a qualitative understanding of the circumstances under which this type of switching will occur.  A positive voltage applied to the gate will lower the energy levels of the molecule, and a negative voltage will raise the energy levels.  The energy levels of the source and drain electrodes are unaffected by the application of the gate voltage, so the equilibrium chemical potential of the contacts, $\mu$, will not change.  As we apply various voltages to the gate electrode, the energy levels of the molecule will reside at different positions relative to the chemical potential of the unbiased source and drain electrodes (Figure 4).  By definition, a molecular energy level residing below the Fermi level is filled, and an energy level above the Fermi level is empty.  Non-zero conduction measured immediately outside of the conductance gap is a result of current flowing through the first energy level of the molecule that is caught between the biased source and drain electrochemical potentials.  Since we are assuming that $\mu_1$ and $\mu_2$ offset more or less equally, we can then argue that conduction will first occur through whichever molecular energy level is closest to the Fermi level, regardless of whether it is above or below $\mu$ at equilibrium.  At zero source-drain bias and zero gate bias, the Fermi level of the system does not lie halfway between the HOMO and LUMO levels, but rather lies\cite{Ghosh2002,Kergueris1999} closer to the HOMO level (Figure 4b).  This implies that for $V_g = 0$, both positive and negative biasing of the source and drain electrodes induces conduction through the HOMO level, and therefore results in the flow of holes (i.e. p-type conduction).  As $V_g$ is decreased from zero, and the HOMO level is pulled upward, we eventually reach the critical point, $V_g = V_C$, at which point the charge state of the molecule is changed by one electron (i.e. one electron is removed from the molecule).  When $V_g = V_C$ the equilibrium Fermi level of the electrodes is aligned with the HOMO level of the molecule, and the conductance gap all but disappears, since any nonzero voltage applied to the source and drain will cause the HOMO level to reside between the offset electrochemical potentials of $\mu_1$ and $\mu_2$.  After passing this critical point, $V_C$, by continuing to decrease $V_g$, what was the HOMO level is now above the Fermi level, and is therefore now unoccupied (Figure 4c), since one electron has been removed.  The result is that when the source and drain electrodes are biased we measure conduction through a now empty state, which is to say that the current arises from the flow of electrons (i.e. n-type conduction).  As stated before, conduction will first occur through whichever energy level of the molecule is closest to $\mu$, hence we will observe n-type conduction persist, as in Figure 4c, through the same energy level as we decrease $V_g$ until a different molecular energy level becomes the closest energy level to $\mu$.  This condition occurs when $V_g = V_T$, at which point the Fermi level reaches halfway between the energy level that conduction is presently occurring through and the adjacent level down.  By further decreasing $V_g$ past this point, we find that current is now flowing through this next energy level down, which is now the closest energy level to $\mu$ (Figure 4d.).  Because this new energy level is below the Fermi level of the electrodes, we return to the scenario of p-type conduction, or current flowing through a filled energy level.  If we were to have started at $V_g$ = 0 and increased the gate voltage, instead of decreasing it, we would have observed hole conduction persist until we reached $V_g = V_{T'}$.  Increasing $V_g$ past this point, we would observe current flow through the next level up, since that upper level would now be the closest energy level to $\mu$(Figure 4e).  Conduction would now be occurring through an energy level above the Fermi level, hence the system would have switched to n-type conduction. 

In our experiments to date, we have been limited to the addition or subtraction of a single electron to the molecule in question, which means that the source-drain voltage range of our stability diagrams has been restricted.  Currently, the molecules we have been using have no means of forming chemical bonds to the gold source and drain electrodes.  This deficiency makes them less than perfectly stable as the energy of the system is increased.  As a larger $V_g$ is applied, the molecule is prone to conformal changes as well as thermal vibrations and will frequently depart from its position in the breakjunction gap thereby destroying the device.  If we were able to continue increasing or decreasing $V_g$ while still obtaining meaningful data, we would expect to observe a stability diagram with repeated diamond patterns from the opening and closing of the conductance gap.  We would also expect to see differential conductance symmetry reversals (exchanges of $\partial{I}/\partial{V}$ peaks at positive and negative $V_{sd}$) at every half of a diamond pattern.  In other words, we predict a switch between p-type and n-type conduction at every point like $V_g$, where the conductance gap vanishes, and at every point like $V_T$, where the conductance gap reaches a maximum.  Considering the molecule-electrode arrangement of Figure 1e, we can use the semiclassical model to describe the SMT in terms of a circuit composed of a molecule separated from the source and drain electrodes by two tunnel junctions with capacitances $C_1$ and $C_2$, respectively.  The gate is also coupled to the molecule with capacitance $C_g$.  $E_{Coul}$ is the Coulomb charging energy of the molecule, and $E_{G,n}$ is the gap between adjacent molecular energy levels $E_n$ and $E_{n+1}$.  For this configuration, an exchange between n-type and p-type conduction will always occur when 

\begin{equation}
V_g = \frac{C_{eq}}{2eC_g} (E_{G,n} + E_{Coul}),
\end{equation}

where $C_{eq} = C_1 + C_2 + C_1$.
\vspace{0.2cm}\\

\textbf{CONCLUSION} Utilizing the electromigration technique, we have fabricated single molecule transistors using a Borromean Ring complex as the active element.  The devices were measured as functions of both source-drain voltage and gate voltage.  We observe a reversal of the asymmetric nature of the stability diagrams that is manifested most dramatically as an interchange between differential conductance peaks on opposing sides of the conductance gap.  This process is exhibited as a function of gate bias.  We interpret this phenomenon as the result of a switch between p-type and n-type conduction, dependent upon the position of the electrode equilibrium Fermi level, with respect to the HOMO and LUMO levels of the transistor molecule for a given gate voltage.

The ability to exert further control over novel nanometer scale devices, such as SMTs, boosts the potential for their eventual integration into mainstream electronic devices.  The characteristic conduction switching in SMTs described here is a general property of SMTs and not unique to the Borromean Ring complexes.  Details of SMT transport, such as those we report here, lead us toward a larger picture of conduction through single molecules and other nanoscale devices, furthering the case for SMTs as molecular probes and leading ultimately to a more widespread use of molecules in electronics.

This work is supported by the Defense MicroElectronics Activity  (DMEA 90-02-2-0217) and by the National Science Foundation (CHE0317170).

\end{document}